\documentclass[aps,prb,twocolumn,final,groupedaddress]{revtex4}

\usepackage[intlimits]{amsmath}
\usepackage{amssymb}
\usepackage[draft]{graphicx}
\usepackage{graphicx}
\usepackage{braket}
\usepackage{color}
\definecolor{MidnightBlue}{RGB}{0,0,160}
\usepackage[pdftitle={AHE in disordered ferromagnets}, pdfauthor={Philippe Czaja}, colorlinks=true, citecolor=MidnightBlue,linkcolor=black,urlcolor=MidnightBlue, pdftex, breaklinks=true]{hyperref}

\newcommand{\ve}[1]{\mathbf{#1}}

\newcommand{\un}[1]{\text{ #1}}
\newcommand{\im}[1]{\Im{\left\lbrace #1 \right\rbrace}}
\newcommand{\re}[1]{\Re{\left\lbrace #1 \right\rbrace}}

\newcommand{\md}[1]{\mathrm d #1 }
\newcommand{\me}{\mathrm e}

\newcommand{\Tr}[1]{\mathrm{Tr}\left[ #1 \right]}
\newcommand{\mtr}[2]{\mathrm{Tr}\left[ #1 \right. \\ \left. #2 \right]}

\newcommand{\del}[2]{\frac{\partial #1}{\partial #2}}

\newcommand{\diac}[2]{\raisebox{#1\dp\strutbox}{\raisebox{-.5\height}{\includegraphics[scale = 0.46]{diagrams/#2.pdf}}}}

\begin{document}

\title{Anomalous Hall effect in ferromagnets with Gaussian disorder}

\author{Philippe Czaja}
\email[]{p.czaja@fz-juelich.de}
\author{Frank Freimuth}
\author{J\"urgen Weischenberg}
\author{Stefan Bl\"ugel}
\author{Yuriy Mokrousov}
\email[]{y.mokrousov@fz-juelich.de}
\affiliation{Peter Gr\"unberg Institut and Institute for Advanced Simulations, Forschungszentrum J\"ulich, 52425 J\"ulich, Germany}

\date{\today}

\begin{abstract}
Using the Kubo formalism we derived expressions and implemented the method for calculating the anomalous Hall conductivity (AHC) in ferromagnets with short-range Gaussian disorder directly from first-principles electronic structure of the perfect crystal. We used this method to calculate the AHC  in bcc Fe, fcc Co, L1$_0$-FePd, L1$_0$-FePt as well as thin bcc Fe(001) films. Within our approach we can transparently decompose the conductivity into intrinsic (IC), side jump (SJ) and {\it intrinsic skew-scattering} (ISK) contributions. The existence of ISK, which originates from asymmetric Mott scattering but is clearly distinguishable from conventional skew-scattering in that it converges to a finite value in clean limit, was pointed out by 
Sinitsyn {\it et al.} [Phys. Rev. B {\bf 75}, 045315 (2007)]. Here, we collect
{\it all} contributions to the AHC in ferromagnets which result in ``scattering-independent" AHE in clean limit, and analyze their
relative magnitude from first principles calculations. By comparing our results to existing experiments we show that the Gaussian disorder is well suited to model various types of disorder present in real materials, to some extent including the effect of temperature. 
In particular, we show that in addition to intrinsic and side-jump AHE, the intrinsic skew-scattering can be a major player in determining the magnitude of the AHE
in ferromagnets.
\end{abstract}


\maketitle

\section{Introduction \label{sec:introduction}}
The anomalous Hall effect (AHE)\cite{Hall} has been known and experimentally investigated for a long time, but until recently has eluded a rigorous quantum mechanical description. The revival of interest in the AHE during the last
few years was mainly driven by the discovery of new applications in spintronics, by the new interpretation
of electronic properties from the point of view of the Berry phase theory,\cite{Karplus,Nagaosa06} and also by the
increasing computational resources which made possible the computationally very expensive {\it ab initio} assessment
of the AHE in transition-metals.\cite{Nagaosa}

The reason why the theoretical description of the AHE, and even more its calculation, is so challenging is that unlike the ordinary Hall effect, which can be intuitively explained in terms of the Lorentz force, the AHE originates from a variety of elaborated physical mechanisms. Within the semi-classical theory, three different contributions to the AHE can be distinguished according
to the underlying processes:\cite{Nagaosa} the {\it intrinsic} contribution, which is the topological property of the band structure and exists even without disorder, and two disorder-driven {\it extrinsic} contributions. These are the {\it skew scattering},\cite{Smit58} which arises from an asymmetric Mott scattering of electrons off impurities in the presence of spin-orbit interaction, and the {\it side jump},\cite{Berger70} which can be formally defined as the remaining part of extrinsic scattering and was historically interpreted as a sideways displacement of electrons scattering off impurities when spin-orbit interaction is present.

Experimentally, distinguishing different contributions from their scaling behavior 
with the impurity concentration and temperature does not present a trivial task.\cite{Dheer,Smit55,Koetzler,Tian,Ye} 
In this context \emph{ab initio} calculations of the AHE play an important role since in principle they allow us to access different contributions separately and explicitly investigate the effect of disorder, thus presenting an important tool for understanding and engineering the behavior of the AHE in real materials.
Previous \emph{ab initio} calculations of the AHE have mainly focused on the IC, for which an explicit expression in terms of the Berry curvature of the perfect crystal has been known for some time.\cite{Nagaosa} Calculations of the AHE 
including disorder due to
alloying or an alloy-analogy model for a set of thermal lattice displacements have been performed using the Korringa-Kohn-Rostoker (KKR) method 
in combination with the coherent potential approximation (CPA).\cite{Lowitzer,Turek,Koedderitzsch} 
These methods rely on exact knowledge of the disorder potential, which has to be explicitly included in the \emph{ab initio} calculation. This makes the calculation of the disorder driven contributions by far more complicated and expensive than the calculation of the intrinsic part.

With vanishing disorder, a certain part of the AHE, which we call the {\it scattering-independent} AHE, acquires a constant value,
which is believed to be disorder-independent.\cite{Nagaosa} Treating disorder within a short-ranged Gaussian disorder model, this scattering-independent contribution to AHE can be
identified in the clean limit.\cite{Nagaosa} The scattering-independent AHE can be conveniently calculated from the electronic structure of the pure crystal
even for multi-band metals\cite{Kovalev} and provides the dominant source of the AHE in transition-metal ferromagnets which 
are moderately disordered.\cite{Weischenberg}  In a sense, the clean limit scattering-independent AHE 
presents the ground level value, around which the disorder-sensitive contributions arise. The combination of the
short-ranged Gaussian disorder model with realistic electronic structure calculations is thus a very
attractive alternative approach to treating the effect of disorder on the AHE and related phenomena on the {\it ab initio} 
level.

In the past, based on the semiclassical Boltzmann equation and Kubo-Streda formalism, it was argued by Sinitsyn and co-workers\cite{Sinitsyn-2,Sinitsyn} that an additional to IC and SJ scattering-independent contribution to the AHE is provided by the so-called {\it intrinsic
skew-scattering}. As the conventional skew-scattering, which is inversely proportional to the impurity concentration and thus
arises due to incoherent superposition of scattering at each defect, ISK
also originates from asymmetric part of the collision kernel, but it reflects the interference between scattering at different 
impurities and it scales the same as SJ and IC. Diagrammatically speaking,
while the conventional skew-scattering is due to the vertex corrections that involve correlators of more
than two powers of the disorder potential, the ISK is only due to Gaussian disorder correlations.\cite{Sinitsyn} The importance of ISK in terms of its relation
to other contributions on a model
level or in ferromagnetic materials was never investigated.

In this work, based on the formalism of Kovalev and Weischenberg,\cite{Kovalev,Weischenberg} we derive analytical expressions for the AHE in 
the presence of finite Gaussian disorder which can be evaluated solely from the electronic structure of the perfect crystal. This is done by constructing the unperturbed Green functions based on a preceeding electronic structure calculation, and then applying the Gaussian disorder model to obtain the self-energy, from which the full Green function and the vertex corrections can be calculated. In
the clean limit we thus obtain {\it all} scattering-independent contributions, which, in addition to IC and SJ considered 
previously,\cite{Weischenberg} also include the intrinsic skew-scattering. We implement derived expressions in the first-principles
full-potential linearized augmented plane-wave (FLAPW) code {\tt FLEUR},\cite{FLEUR} and calculate the AHE in the clean limit and away from it in typical metallic ferromagnets studied theoretically
and experimentally: bcc Fe, fcc Co, L1$_0$-FePd, L1$_0$-FePt and thin films of bcc Fe(001).

Overall, we obtain a good agreement of our results with experimental data, indicating that the Gaussian disorder model is well suited for modeling the effect of impurities and to some extent even effect of temperature. Analyzing the individual contributions,
 unambiguously distinguishable within our approach, we find that in most materials neither intrinsic nor extrinsic AHC shows a trivial behavior as a function of resistivity or temperature. One of our key findings is that ISK is just as important as the IC and SJ, and including it into consideration systematically improves agreement with experiment. We can also conclude that the main contribution to the AHE, which completely drives its behavior with respect to disorder in the investigated materials, comes from the Fermi surface, whereas the Fermi sea provides a smaller contribution that is quite insensitive to disorder.

\section{Method \label{sec:method}}
We can use the Kubo formalism to derive the expression for the anomalous Hall conductivity (AHC) at zero temperature in terms of the retarded and advanced Green functions $G^{R/A}_0$ of the perfectly periodic crystal and the velocity operator $\ve v$. This gives rise to a sum of two terms (throughout this section we set $e=\hbar=1$),
\begin{equation}
\begin{aligned}
\sigma_{\alpha\beta}^{\mathrm{I}} = \frac{1}{4\pi}\int \frac{d^3  k}{(2\pi)^3}
\mtr{v^\alpha G_0^R(E_F,\ve k) v^\beta G_0^A(E_F, \ve k)}{- (\alpha \leftrightarrow \beta)}
\end{aligned}
\label{eq:sigma_1}
\end{equation}
and
\begin{equation}
\begin{aligned}
&\sigma_{\alpha\beta}^{\mathrm{II}}= \frac{1}{2\pi} \int \frac{d^3 k}{(2\pi)^3} \int_{-\infty}^{E_F}\md{E}\times\\ 
&\qquad\re{ \Tr{v^\alpha G^R_0(E,\ve k)^2 v^\beta G^R_0(E, \ve k) - (\alpha \leftrightarrow \beta)}} ~,
\end{aligned}
\label{eq:sigma_2}
\end{equation}
where $\alpha$ and $\beta$ are the cartesian indices and $E_F$ is the Fermi level. $\sigma_{\alpha\beta}^{\mathrm{I}}$ we will refer to as the Fermi-surface term since it has contributions coming only from the Fermi surface. $\sigma_{\alpha\beta}^{\mathrm{II}}$ accumulates the contributions from all occupied states and it is therefore referred to as the Fermi-sea term. Together these two terms yield the anomalous Hall conductivity of a disorder-free crystal. 

In order to treat disordered systems instead of $G_0$ in Eqs.~\eqref{eq:sigma_1} and \eqref{eq:sigma_2} we have to consider the full Green function 
\begin{equation}
G = \frac{1}{(G_0^{-1}-\Sigma)},
\label{eq:Dyson}
\end{equation}
which is obtained from $G_0$ and the self-energy $\Sigma$, incorporating the effect
of disorder. This is done using two different models. Within the first model the self-energy is approximated by a constant imaginary part $\Gamma$. Whereas this model does not treat scattering explicitly and is thus not able to reproduce the disorder driven contributions to the AHE, it is well suited for studying the effect of finite temperatures on the intrinsic contribution. In order to cover also the disorder driven contributions, a short-range Gaussian disorder model together with configurational averaging is used to account for impurity scattering explicitly.

\subsection{Constant broadening}
Setting $\Sigma(E, \ve k)$ to $-i\Gamma \cdot I$, where $I$ is the identity matrix, results in a Green function that is diagonal in the eigenspace of the Hamiltonian:
\begin{equation}
G^{R}(E,\ve k)_{mn} = \frac{\delta_{mn}}{E-\epsilon_{n\ve k}+i\Gamma} ~,
\end{equation}
where $\epsilon_{n\ve k}$ are the single-electron eigen energies. Inserting $G$ instead of $G_0$ into \eqref{eq:sigma_1} and \eqref{eq:sigma_2} respectively yields
\begin{equation}
\begin{aligned}
\sigma_{\alpha\beta}^{\mathrm{I}} = -\frac{1}{2\pi}\int \frac{d^3 k}{(2\pi)^3} \sum_{\underset{m \ne n}{mn}}\im{{v^{\alpha}_{mn}(\ve k)}{v^{\beta}_{nm}(\ve k)}} \times \\
\frac{(\epsilon_{m\ve k}-\epsilon_{n\ve k})\Gamma}{((E_F-\epsilon_{m\ve k})^2+\Gamma^2)((E_F-\epsilon_{n\ve k})^2+\Gamma^2)}
\end{aligned}
\end{equation}
and
\begin{equation}
\begin{aligned}
\sigma_{\alpha\beta}^{\mathrm{II}} &= \frac{1}{\pi}\int \frac{d^3 k}{(2\pi)^3} \sum_{\underset{m \ne n}{mn}}\im{{v^{\alpha}_{mn}(\ve k)}{v^{\beta}_{nm}(\ve k)}} \times \\
&\left(\frac{\Gamma}{(\epsilon_{m\ve k}-\epsilon_{n\ve k})((E_F-\epsilon_{m\ve k})^2+\Gamma^2)} \right. \\
&\left. -\frac{1}{(\epsilon_{m\ve k}-\epsilon_{n\ve k})^2}\im{\ln{\frac{E_F-\epsilon_{m\ve k}+i\Gamma}{E_F-\epsilon_{n\ve k}+i\Gamma}}}\right) ~,
\end{aligned}
\end{equation}
where all matrices are given in the eigenbasis of $H$. In the clean limit $\Gamma \to 0$ the sum of both terms converges to the well-known Berry curvature expression for the intrinsic AHE,
\begin{equation}
\sigma_{\alpha\beta} \to \sum_m^\text{occ} \int \frac{d^3 k}{(2\pi)^3} \sum_{n\ne m} \frac{2\im{{v^{\alpha}_{mn}(\ve k)}{v^{\beta}_{nm}(\ve k)}} }{(\epsilon_{m\ve k}-\epsilon_{n\ve k})^2} \label{eq:sigma_int} ~.
\end{equation}

\subsection{Gaussian disorder}
In order to derive the scattering-originated contributions to the AHE we use the model potential 
\begin{equation}
\hat V = U \sum_{i}^{N_\text{imp}} \delta(\hat{\ve r}-\ve R_i), 
\end{equation}
which consists of a number $N_\text{imp}$ of delta functions at positions $\ve R_i$.
 We calculate the averaged Green function $G_\text{av} = \braket{G}_c$, which is obtained by taking the configurational average over all possible distributions of $N_\text{imp}$ impurities. In the following we will refer to $G_\text{av}$ as $G$ and treat it  as a regular Green function which fulfills the Dyson equation \eqref{eq:Dyson}. In order to obtain an expression for the self-energy $\Sigma$, we expand it in powers of the potential $V$ and perform the configurational average, which yields an infinite series of diagrams, each describing a multiple scattering event:
\begin{equation}
\diac{.8}{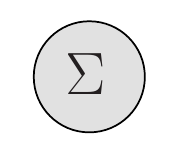} = \diac{.8}{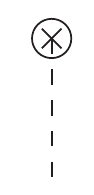} +\diac{.8}{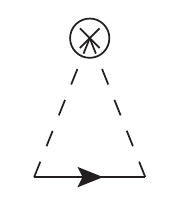} +\diac{.8}{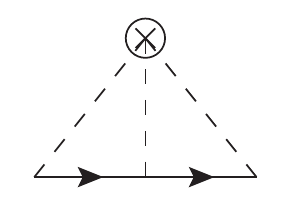} +\diac{.8}{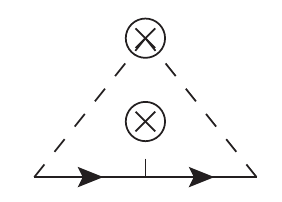} +\hdots
\end{equation}
Here, a single line represents a disorder-free Green function $G_0$ and a scattering amplitude is represented by a dashed line and a cross, where the cross stands for an impurity.
The first order term only gives a constant energy shift that can be hidden in the chemical potential and is thus irrelevant. Truncating the series after the first non-trivial term yields
\begin{equation}
\Sigma(E,\ve k) = \mathcal{V}\int \frac{d^3 k'}{(2\pi)^3}O_{\ve k\ve k'}G_0(E,\ve k')O_{\ve k'\ve k} ~,
\label{eq:Sigma}
\end{equation}
where $\mathcal V = U^2n_\text{imp}$ is a disorder parameter containing the disorder strength $U$ and the disorder concentration $n_\text{imp}$, and $O_{\ve k,\ve k'}$ are the overlap matrices of the lattice periodic parts of the Bloch states: $(O_{\ve k,\ve k'})_{mn} = \braket{\ve k m | \ve k' n}$. This truncation is justified by the fact that the self-consistent evaluation of \eqref{eq:Sigma} reproduces all higher order terms which are important at low impurity concentrations.

For a complete description of the AHE we also need to include the vertex corrections,~i.e., all the terms which are not obtained when the full Green function is inserted into \eqref{eq:sigma_1} and \eqref{eq:sigma_2}. Which terms these are can be made clear diagrammatically. In terms of Feynman diagrams the Fermi surface term is represented as a bubble
\begin{equation}
\diac{.8}{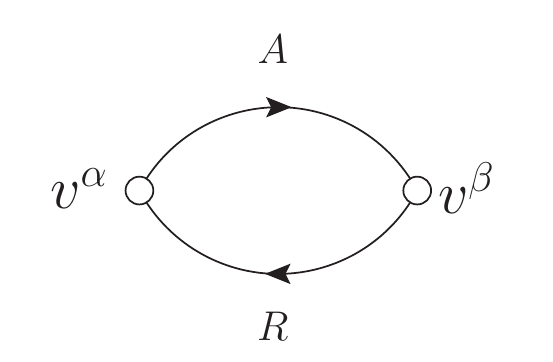}
\end{equation}
where open circles represent velocity vertices. When $G_0$ is replaced by $G$, represented by a double line, we obtain
\begin{equation}
\diac{.8}{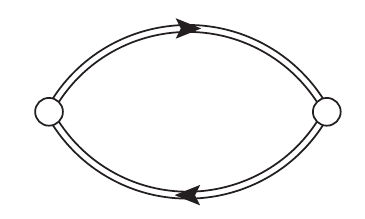} = \diac{.8}{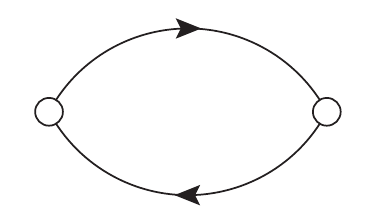} + \diac{.8}{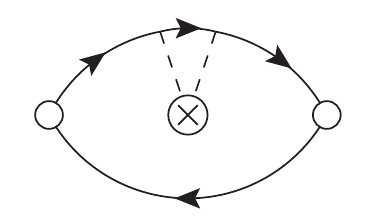} + \diac{.8}{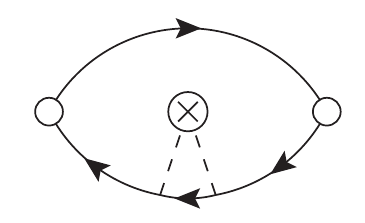} + \hdots
\end{equation}
This series does not contain processes where the top and the bottom Green function are connected by one or more scattering lines. These processes are however equally important and therefore have to be taken into account, which is done by replacing the velocity vertex $\ve v$ by a vertex function $\ve \Gamma(E, \ve k)$ (represented by a gray triangle). Again neglecting processes with more than two scatterings from the same impurity, we find that the vertex function is given by an infinite series of so-called ladder diagrams:
\begin{equation}
\diac{.8}{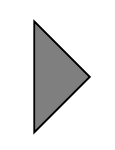} = ~\diac{.8}{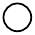}~ + \diac{.8}{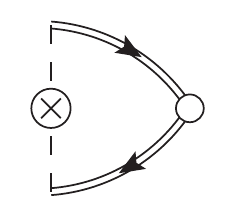} + \diac{.8}{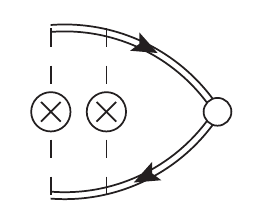} + \hdots
\end{equation}
This can be written in form of a self-consistent equation,
\begin{equation}
\diac{.8}{Gamma} = ~\diac{.8}{vvert}~ + \diac{.8}{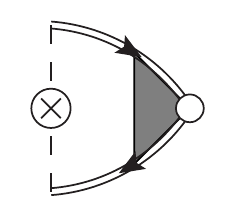} ~,
\end{equation}
or in integral form,
\begin{equation}
\begin{aligned}
\ve \Gamma(E,\ve k) &= \ve v(\ve k) +\\
+&\mathcal{V}\int \frac{d^3 k'}{(2\pi)^3} O_{\ve k\ve k'}G^A(E,\ve k')\ve\Gamma(E,\ve k')G^R(E,\ve k')O_{\ve k'\ve k} \label{eq:Gamma}
\end{aligned}
\end{equation}
which can be solved either iteratively or via matrix inversion. Equally, scalar vertices have to be replaced by a scalar vertex function $\gamma$ for which we obtain a similar equation:
\begin{equation}
\begin{aligned}
\gamma(E,\ve k) &= I +\\
+&\mathcal{V}\int \frac{d^3 k'}{(2\pi)^3} O_{\ve k\ve k'}G^R(E,\ve k')\gamma(E,\ve k')G^R(E,\ve k')O_{\ve k'\ve k}
\label{eq:gamma}
\end{aligned}
\end{equation}
The full AHC is now obtained by replacing in \eqref{eq:sigma_1} and \eqref{eq:sigma_2} $G_0$ by $G$ and the scalar and velocity vertices by the respective vertex functions:
\begin{equation}
\begin{aligned}
\sigma_{\alpha\beta}^{\mathrm{I}} = \frac{1}{4\pi}\int \frac{d^3  k}{(2\pi)^3}
\mtr{\Gamma^\alpha(E_F) G^R(E_F) v^\beta G^A(E_F)}{- (\alpha \leftrightarrow \beta)}
\end{aligned}
\end{equation}
and
\begin{equation}
\begin{aligned}
\sigma_{\alpha\beta}^{\mathrm{II}} &= \frac{1}{2\pi} \int \frac{d^3 k}{(2\pi)^3} \int_{-\infty}^{E_F}\md{E}\times \\ 
&\begin{aligned}\Re \left\lbrace \mathrm{Tr} \left[ \Gamma^\alpha(E) G^R(E) \gamma(E) G^R(E) \Gamma^\beta(E) G^R(E) \right. \right. ~ & \\ 
\left. \left. - (\alpha \leftrightarrow \beta) \right] \right\rbrace & ~,
\end{aligned}
\end{aligned}
\end{equation}
where for clarity the $\ve k$-dependence was omitted for all operators. This leads to the following diagrammatic picture of the AHE, which allows for a clear distinction of its different contributions:
\begin{equation}
\begin{aligned}
\sigma_{\alpha\beta} = &\underbrace{\bigg( \hspace*{-2pt}\diac{.8}{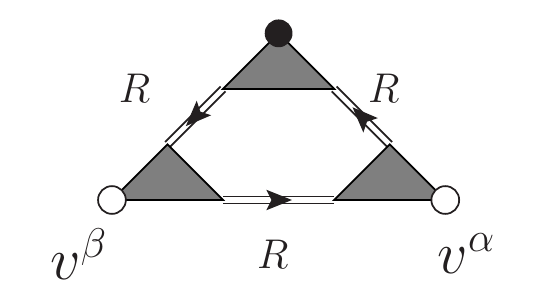} \hspace*{-8pt} - (\alpha \leftrightarrow \beta)\bigg) + \diac{.8}{craab}}_{\sigma_{\alpha\beta}^{int}} \\
&+\underbrace{\diac{.8}{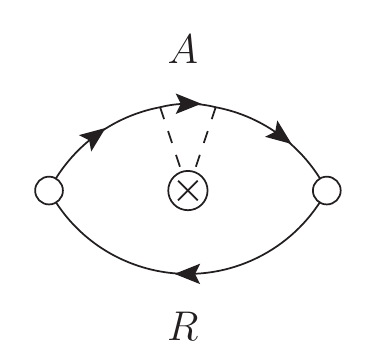} + \diac{.8}{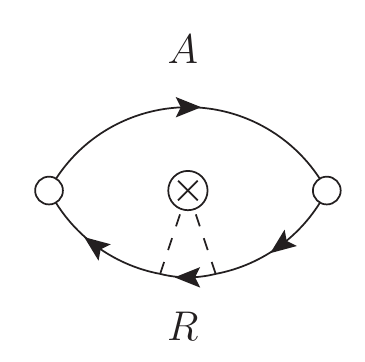} + \hdots}_{\sigma_{\alpha\beta}^{sj}} + \underbrace{\diac{.8}{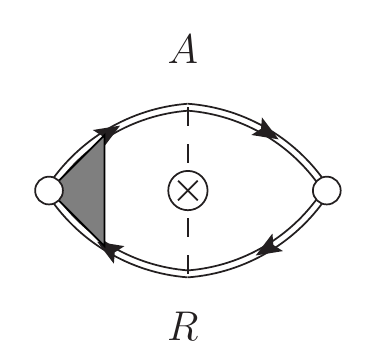}}_{\sigma_{\alpha\beta}^{isk}} \\
&- (R \leftrightarrow A) ~.
\end{aligned} \label{eq:separation}
\end{equation}
The first term is the Fermi sea part, which is usually considered to be of an intrinsic origin\cite{Nagaosa} since it does not show any scattering-driven behavior, i.e., its disorder dependence can be captured by a simple broadening. Together with the disorder-free part of the Fermi surface term it forms the intrinsic AHC 
$\sigma^{int}$, which in the clean limit converges to the Berry curvature expression \eqref{eq:sigma_int}. The side jump $\sigma^{sj}$ is defined by the disorder-driven terms which emerge when replacing $G_0$ by $G$ in the Fermi surface term. In the clean limit it converges to the scattering-independent side-jump contribution.\cite{Weischenberg,Kovalev} The remaining part, which corresponds to the vertex corrections, is called intrinsic skew scattering $\sigma^{isk}$.\cite{Sinitsyn-2,Sinitsyn} Similar to the SJ, it converges to a finite value in the clean limit and thereby differs from the conventional skew scattering. The latter is divergent in the clean limit but it is negligible in the region of impurity concentrations which are usual for moderately 
disordered metals.\cite{Nagaosa} In our work the conventional skew scattering is not considered since it does not arise from the Gaussian disorder model that we use. Speaking diagrammatically, the conventional skew scattering consists of  more elaborated vertex corrections which depend on internal details of impurity potential and which are missing in the representation above.\cite{Sinitsyn,Nagaosa}

\subsection{Implementational aspects and computational details}

In practice the Brillouin zone integrals are replaced by a sum over a discrete $\ve k$-point grid. Since the convergence of the Fermi surface term requires a large number of $\ve k$-points, we use the method of Wannier interpolation \cite{Souza01} to interpolate all the necessary quantities on a sufficiently dense grid. For this purpose we construct a set of maximally localized Wannier functions (MLWF) $\ket{\ve Rn} = 1/\sqrt{N_k}\sum_{\ve k}\me^{-i\ve k\cdot \ve R}\ket{\psi_{n\ve k}^W} $ on a coarse $\ve k$-point grid, where $\ket{\psi_{n\ve k}^W}=\sum_m U_{mn}^{\ve k} \ket{\psi_{m\ve k}}$ are the Bloch states in the Wannier gauge, obtained by a unitary mixing of the eigenstates $\ket{\psi_{m\ve k}}$. We then calculate the Hamiltonian $H(\ve R)$ in the Wannier basis and perform an inverse Fourier transformation to obtain the Hamiltonian $H^W(\ve k)$ in the Wannier gauge,~i.e., in the basis $\{\ket{\psi_{n\ve k}^W}\}$. The matrix elements of the velocity operator are then obtained according to\begin{equation}{\ve v(\ve k)}^W_{mn} =\left(\del{H(\ve k)^W}{\ve k}\right)_{mn} \approx \del{H(\ve k)^W_{mn}}{\ve k} ~,\label{eq:vk} \end{equation}where the approximation made in \eqref{eq:vk} becomes valid by assuming that $ \partial_{\ve k} \ket{\ve k n}^W \approx 0$,  where $\ket{\ve k n}^W$ is the lattice periodic part of the Bloch wavefunction in the Wannier gauge  (constant basis approximation).
For the evaluation of the AHC in the constant broadening model both $H$ and $\ve v$ are rotated into 
the eigenbasis of $H$, whereas in the Gaussian disorder model all matrices are evaluated in the Wannier
 gauge. Due to the constant basis approximation the overlap matrices in the Wannier gauge simplify to unit matrices.
 The only input quantity remaining is therefore the Hamiltonian in the Wannier basis. The energy integrals in the Gaussian disorder model are evaluated numerically using a complex energy grid \cite{Wildberger} with 15 or 31 points.

The electronic structure calculations were carried out within the full-potential linearized augmented plane-wave method using the J\"ulich density-functional theory FLAPW code {\tt FLEUR}.\cite{FLEUR} For bulk calculations we used experimental lattice constants and between $7700$ and $9300$ $\ve k$-points in the full Brillouin zone together with a planewave cut-off $k_\text{max}$ between $3.7$ and $4.5 \un{bohr}^{-1}$ to ensure convergence of the charge density. In all cases
the ferromagnetic magnetization was pointing along the [001] axis. The construction of Wannier orbitals was performed with the \textsc{Wannier90} code \cite{wannier90} and our interface between {\tt FLEUR} and \textsc{Wannier90}.\cite{interface} In all bulk calculations we constructed a set of $18$ MLWFs per atom on an $8\times 8\times 8$ grid using $d_{xy}$-, $d_{xz}$-, $d_{yz}$- and $sp^3d^2$-orbitals as first guesses. 
For the film calculations we used the 2D version of {\tt FLEUR}\cite{FLEUR} with a distance of $a/2=2.71\un{bohr}$ between the Fe layers, $5.1\un{bohr}$ between the surface layer and the vacuum boundary, and $6.8\un{bohr}$ between the surface layer and the $z$-boundary used to generate the plane-waves along the $z$-axis,
normal to the film plane. All film calculations were performed with $784$ $\ve k$-points in the full Brillouin zone and $k_\text{max} = 3.8 \un{bohr}^{-1}$. For the 3-layer film we constructed $18$ MLWFs  per layer using the same first-guess orbitals as in the bulk case. For the other 
films we constructed only $12$ MLWFs per layer, starting from $s$- and $d$-orbitals. In all cases an $
8\times 8$ grid was used. In order to calculate the AHE in a film the expressions in the previous sections can be straightforwardly adapted to two dimensions, with the only difference that the conductance is calculated instead of the conductivity. The AHC is then obtained from dividing by the film thickness, which was approximated by $(N_\text{layers}+1) a/2$.

\section{Results \label{sec:results}}

The method which we introduced above was used to calculate the AHE in disordered ferromagnets solely based on the \textit{ab initio} electronic structure of the perfect crystal. This approach has already been successfully used by Kovalev \cite{Kovalev} and Weischenberg \cite{Weischenberg} to calculate the SJ in clean limit. In this work we extend the latter approach by introducing finite disorder in the form of a single disorder parameter into the calculation. This has the advantage that a systematic treatment and an easy tuning of disorder, which enables the straightforward application of the method to any new material, is possible. Moreover, our approach allows us to unambiguously separate the AHE into different contributions according to \eqref{eq:separation}. This could
be particularly beneficial for comparison with the model studies\cite{Nagaosa} and for establishing
the AHE scaling laws relevant experimentally.\cite{Tian,Ye,Hou,He,Seemann} In particular, 
this allows us to investigate the importance of the intrinsic skew scattering contribution in ferromagnets both at finite disorder and in the clean limit, which has not been considered so far.

  \begin{figure*}
 \includegraphics[scale = 0.97]{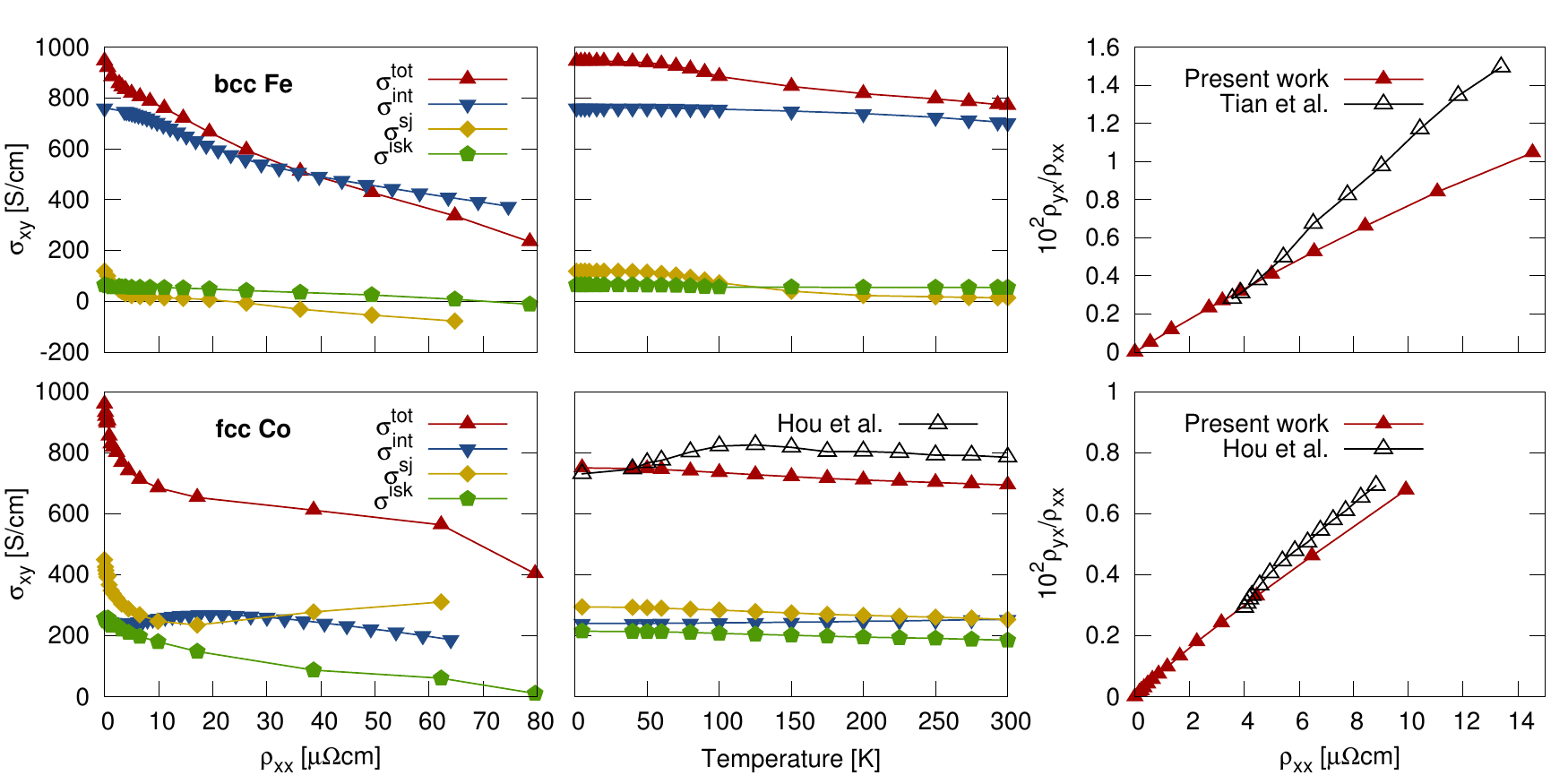}
 \caption{AHC in bcc Fe (top row) and fcc Co (bottom row). Left and middle panels: 
additive  decomposition of the total AHC ($\sigma^{tot}$), into intrinsic ($\sigma^{int}$), side jump ($\sigma^{sj}$) and intrinsic skew scattering ($\sigma^{isk}$) contributions, presented as a function of longitudinal resistivity (left) and temperature (middle). Right panel: $\rho_{yx}/\rho_{xx}$ versus $\rho_{xx}$. Labels "Tian {\it et al.}" and "Hou {\it et al.}" refer to experimental data
in Refs.~[\onlinecite{Tian}] and [\onlinecite{Hou}], respectively.
\label{fig:1}}
 \end{figure*}
 
For the calculation of the IC the effect of disorder was simulated by broadening of the bands, as described in section \ref{sec:method}. 
 Whereas the ISK is calcualted directly, the SJ results from the difference between the total AHC without vertex corrections and the IC.  In order to be able to compare our results at finite disorder to experimental values it is necessary to relate the abstract disorder parameter to an experimentally accessible quantity. This was done by additionally calculating the diagonal conductivity, which allows to plot $\sigma_{\alpha\beta}$ as a function of the longitudinal resistivity $\rho_{\alpha\alpha}$. 
 
 It is clear that in reality the AHC as well as the resistivity depend on a variety of factors such as the impurity concentration, the type of disorder and its microscopic details. Thus, the description in terms of a single parameter describing disorder can be only a rough approximation. In this respect, when the objective is to simulate the effect of a particular source of disorder, our approach is inferior to methods where disorder is explicitly considered in the \textit{ab initio} calculation, such as KKR plus CPA approach.\cite{Koedderitzsch, Lowitzer,Turek} However it seems more general in the sense that it takes into account all kinds of disorder sources and bundles them in an average disorder potential. In how far this is a reasonable description of the experimental situation, where one usually also deals with a variety of (often unknown) sorts of disorder acting together, and in how far our method is able to simulate the effect of particular sources of disorder, can be found out only via comparison to experimental data. 
 
 One of the objectives of this work is tackling the question of how well the Gaussian disorder model is capable of modeling the effect of temperature on the AHE. For this purpose we use the values of experimental resistivity as a function of temperature to derive the temperature dependence of the AHC from its resistivity dependence. The results are then compared to experiments.
One has to note that although the temperature dependence of the AHC can come from various sources,  probably the most important source of scattering which our approach is meant to reproduce is the scattering at phonons, since we compare our calculations to experiments performed quite far from
 the Curie temperature for considered materials.

 The temperature as well as the resistivity dependence of the AHC and its individual contributions will now be discussed separately for each investigated material.

\subsection{bcc Fe}

The results of our calculations of the AHE in bcc Fe are presented in the upper panel of Fig.~\ref{fig:1}.
As we can see, the AHE in Fe is largely driven by the intrinsic contribution $\sigma^{int}$ over a wide range of diagonal resistivity $\rho_{xx}$. The dependence of the AHC on $\rho_{xx}$ is quite pronounced, with an overall reduction by a factor of 3 as $\rho_{xx}$ reaches 80$\mu\Omega$cm. The SJ part $\sigma^{sj}$ is much smaller when compared to the intrinsic conductivity values, which is in accordance with Weischenberg's clean limit calculation for Fe.\cite{Weischenberg} The ISK contribution to the AHC, $\sigma^{isk}$, is also small, although is becomes dominant over side jump in the vicinity of the $\rho_{xx}$
where side jump changes sign. Both extrinsic contributions do not exceed 150 S/cm in absolute value
for the whole considered resistivity range. In the clean limit, adding all contributions to the AHC results 
in the value of about 950~S/cm, which is very close to the value of about 1000$-$1100~S/cm for the  ``intrinsic", or, as
it should be properly addressed, scattering-independent AHC in bcc Fe, obtained from
carefully crafted recent measurements by Tian {\it et al.}\cite{Tian} In this respect, the inclusion of the intrinsic skew-scattering improves the agreement between theory and experiment.\cite{Weischenberg}

Converting the $\rho_{xx}$ dependence into a temperature ($T$) dependence using experimental resistivity data from Ref.~[\onlinecite{Ho}], we obtain a relatively small decrease of the AHC over temperature leading to a room temperature value of 
800~S/cm.
The decrease of the AHC by about 20\% as compared to zero $T$ can be solely attributed to a decrease in $\sigma^{sj}$, whereas the IC and ISK contributions remain basically constant over $T$ below 300~K. Noticably, at 300~K the  scattering is dominated by intrinsic skew scattering. 

Since experimental values for the AHC as a function of temperature are not available for iron, we instead plot $\rho_{yx}/\rho_{xx}$ versus $\rho_{xx}$, where $\rho_{yx} \approx \sigma_{xy}\rho_{xx}^2$ is the transverse resistivity. In this representation we can compare our results to experimental data by Tian {\it et al.} in Ref.~[\onlinecite{Tian}], who measured $\rho_{yx}$ and $\rho_{xx}$ at varying temperatures, see right panel of Fig.~\ref{fig:1}. Qualitatively, the calculated values of $\rho_{yx}/\rho_{xx}$ show an almost linear behavior that is experimentally also observed in iron at higher temperatures, and while theory and experiment agree very well for lower
values of $\rho_{xx}$, at higher $T$ the experimental line has a higher slope than the calculated one. One of the conclusions
we can make from this plot is that the perfomance of the fits and indentification of extracted parameters with different sources
of the AHE has to be done with care, especially when the $\rho_{xx}$ is tuned by varying $T$.~E.g.~in bcc Fe the slope of the linear
fit of the theoretical data in the above plot gives the value of 730~S/cm, which is even lower than the room temperature value of the
calculated AHC. The corresponding fit of the experimental data yields the value of 1280~S/cm, also obviously off the experimental value for the scattering-independent AHC in Fe.
 
In section \ref{sec:method} we have decomposed the AHC into Fermi sea and Fermi surface terms which can be calculated separately. The calculation of these two terms in bcc Fe yields that the behavior of the AHC as a function of resistivity is purely driven by the Fermi surface term. This is in accordance
to the observation that the main origin of the variation of the AHC with $\rho_{xx}$ is the side jump. The Fermi sea contribution on the other hand varies very slowly as a function of $\rho_{xx}$, but, unlike in other materials considered in the following, is not negligible in Fe, where it contributes approximately $20 - 30 \%$ to the total AHC.

   \begin{figure*}
 \includegraphics[scale = 0.97]{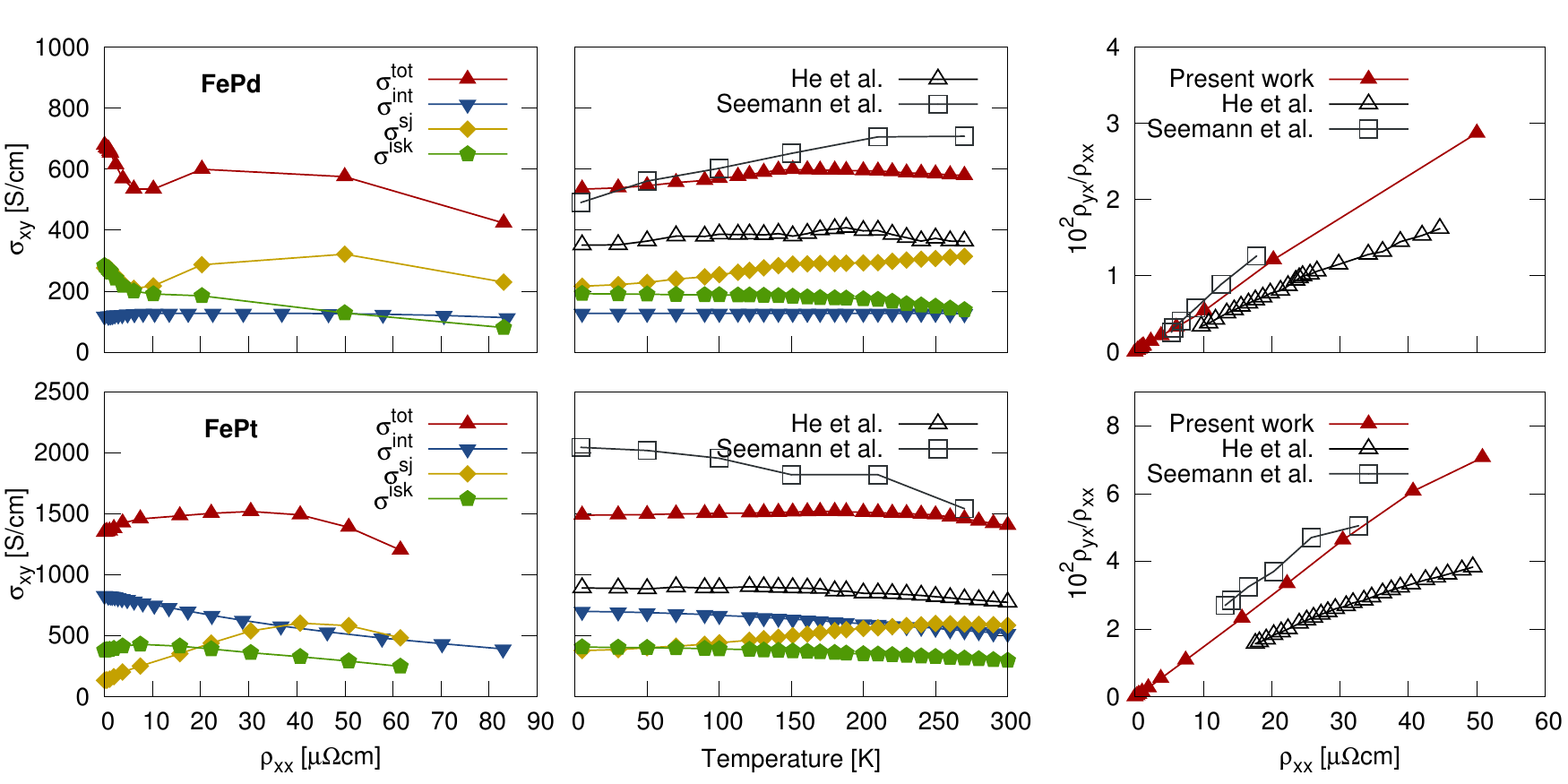}
 \caption{AHC in FePd (top) and FePt (bottom). Presentation of data is analogous to Fig.~\ref{fig:1}. Labels "He {\it et al.}" and "Seemann {\it et al.}" refer to experimental data
in Refs.~[\onlinecite{He}] and [\onlinecite{Seemann}], respectively.\label{fig:2}}
 \end{figure*}

\subsection{fcc Co}

The AHC in fcc Co shows a behavior close to that in bcc Fe, see lower panel of Fig.~\ref{fig:1}, in that
it exhibits similar decay rate with $\rho_{xx}$. Also the  clean limit values of total AHC are very close in both materials. However, unlike in iron, all three contributions to the AHC are of the same order of magnitude and it is mostly the SJ and the ISK which are responsible for the
$\rho_{xx}$ dependence, whereas the IC does not vary with $\rho_{xx}$ significantly.
The Fermi sea contribution to the AHC in fcc Co lies well below $10\%$ of the total value and hardly changes with temperature.

In order to compare our calculations to experiment, we use the experimental data for fcc Co by Hou {\it et al.} given in  Ref.~[\onlinecite{Hou}]. We construct the temperature dependence of the AHC using the experimental $\rho_{xx}$ resistivity data taking into account the residual resistivity of $\rho_{xx}(0 \un K) \approx 4 \mu\Omega\un{cm}$. At low temperatures the agreement of our results with experiment is excellent, with the AHC value of about 750~S/cm, which is smaller than the zero-disorder limit value
due to the offset in $\rho_{xx}$ by $\rho_{xx}(0 \un K)$ in resistivity-dependence of the AHC in Fig.~\ref{fig:1} (left).
In Ref.~[\onlinecite{Hou}] the extracted scattering-independent value of the AHC in fcc Co is reported to be 727~S/cm,
and it is compared there to the intrinsic theoretical value of the AHE of about 250~S/cm, reported in Ref.~[\onlinecite{Roman}].
Our calculations show that the agreement between theory and experiment in estimation of the scattering-independent
value of the AHC 
in fcc Co can be improved significantly, if the side jump and the intrinsic skew-scattering are taken into consideration.
At room temperature the deviation between theory and experiment is also quite small, and constitutes about 15\%.
Both in theory and experiment, the AHC is relatively constant with respect to temperature.
As in the case of Fe, the $\rho_{yx}/\rho_{xx}$-dependence in Fig.~\ref{fig:1} (right) displays a linear
behavior, with a very good agreement to experimental data both in the values and the slope.

\subsection{$L1_0$ FePd and FePt}

In the ferromagnetic L1$_0$-ordered alloys FePd and FePt all contributions to the AHE are of equal importance, as can be seen in Fig.~\ref{fig:2}. In both compounds the total AHC is relatively constant
as a function of $\rho_{xx}$, with the values in FePt by a factor of two larger than in FePd. This has
been previously attributed to the different spin-orbit strength of Pt and Pd atoms, which greatly 
influences the IC and SJ contributions, providing a characteristic crossover.
\cite{Weischenberg}
Indeed, in FePt in clean limit $\sigma^{int}$ is by far larger
than the extrinsic contributions, while the situation changes to opposite in FePd.\cite{Seemann,He} In contrast
to previously presented calculations for bcc Fe, the ISK in the clean
limit is very important for these alloys: it is of the same magnitude as the SJ in FePd, and it is much larger
than SJ in FePt, which leads to an underestimation of AHC in both alloys
if $\sigma^{isk}$ is not included.\cite{Seemann,Weischenberg} As the $\rho_{xx}$ is increased, 
the intrinsic skew-scattering in FePd is significantly decreased, while in FePt the SJ grows
considerably in magnitude. As can be seen from Fig.~\ref{fig:2}, 
in FePd the $\rho_{xx}$ dependence is mainly influenced by 
$\sigma^{sj}$, while in FePt it is the competition of the $\rho_{xx}$-dependence of all
contributions which results in a flat total AHC.  In both alloys the Fermi sea contribution remains
relatively unaffected by disorder, being $\approx 60 \un{S/cm}$ in FePd and $\approx 100 \un{S/cm}$ 
in FePt.

In order to reconstruct the $T$-dependence and compare to experiment we use two sets of
available data: measurements of Seemann~{\it et al.}~on highly-ordered alloys in Ref.~[\onlinecite{Seemann}], and data of He {\it et al.}~in Ref.~[\onlinecite{He}] on samples with lower degree of ordering
and smaller film thickness. As follows from our calculations, the total AHC in both alloys is quite
constant up to room temperature. The agreement with the experimental data of Seemann {\it et al.}~on FePd is overall very good, and even excellent for smaller temperatures, which signifies that in this alloy
the AHE is driven mainly by SJ and, importantly, ISK contribution. In agreement with
experiment, the AHC as a function of temperature below 150~K exhibits a slow rise with $T$. The degree of agreement
 is also visible in the $\rho_{yx}/\rho_{xx}$ plot, given on the right of Fig.~\ref{fig:2}. 
 
 On the other hand,
at first sight the agreement with Seemann's values for FePt seems to be much worse, at least at low temperatures. However, here one has to take into account that in these particular samples of FePt 
the skew scattering angle of about 1\% is extremely large, as can be also seen from $\rho_{yx}/\rho_{xx}$
plot, in which extrinsic skew-scattering corresponds to a shift of the whole curve along the $y$-axis. This leads to a conventional skew scattering $\sigma^{exp}_{sk}$ of about  800~S/cm at 4K,
decreasing towards higher temperatures.
Since the conventional skew scattering is not taken into account within our model, it makes sense  to compare theoretical 
values to $\sigma_{tot}^{exp}-\sigma_{sk}^{exp}$, finding that our calculations actually overestimate Seemann's values by 10$-$15\% at lower $T$, thus providing overall a rather good agreement.
Noticably, the slopes of the $\rho_{yx}/\rho_{xx}$ curves are almost identical between theory
and experiment, and the corresponding curves lie very close to each other. 

The agreement with the experimental data of He {\it et al.} is worse, on the
other hand, which we can attribute to the smaller thickness of the samples and their smaller
ordering parameter, the effect of which could result in a systematic shift to lower values by
about 30\% as compared to our theoretical results, also visible in $\rho_{yx}/\rho_{xx}$
dependence. As apparent from comparison of correspoding
XRD spectra,\cite{Seemann-NJP,He} another reason for discrepancy here could be the difference in the structure which
goes beyond the degree of ordering of the samples used
by Seemann {\it et al.}\cite{Seemann} and He {\it et al.}\cite{He}. Nevertheless, one has to note that the overall trend of the AHC with temperature 
is the same from theory as it is in experiments. Moreover, from careful inspection of the curves
for the total AHC in Fig.~\ref{fig:2} for both alloys we can state that even fine features of the
$T$-dependence in experiments of He {\it et al.} are well-reproduced with our calculations $-$~e.g.~the
rise of the AHC up to 150~K for FePd, or the slight decay of the AHC at around
200~K for FePt are clearly noticable in both alloys. This suggests that the Gaussian disorder
model can reproduce the effect of temperature on the AHE in these alloys rather well.
 
\subsection{Fe films}

 \begin{figure}
 \includegraphics[scale = 0.9]{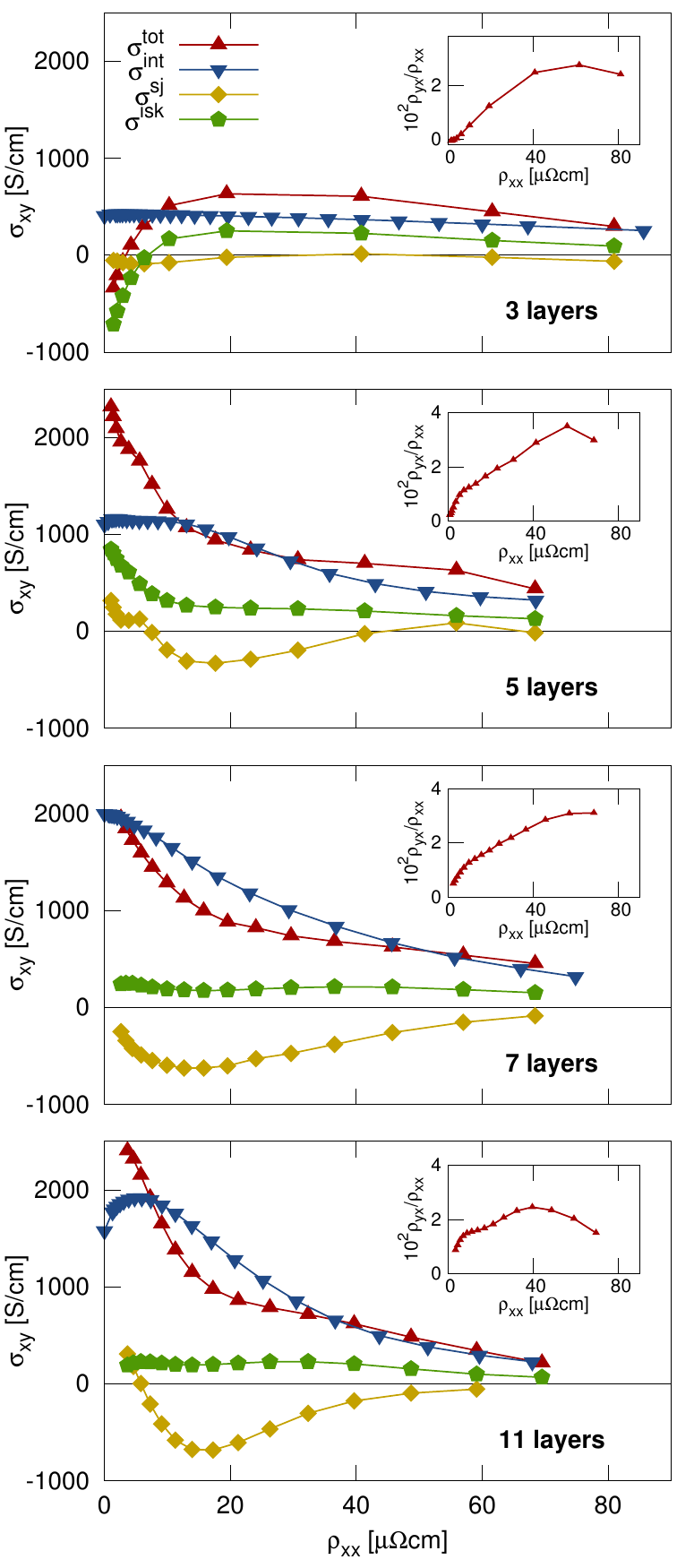}
 \caption{AHC in bcc Fe(001) films of 3-, 5-, 7- and 11-layer thickness. Shown are the total and the decomposed AHC as a function of the longitudinal resistivity. The insets show $\rho_{yx}/\rho_{xx}$ as a function of  $\rho_{xx}$. \label{fig:film}}
\end{figure}

Finally, in addition to bulk Fe we also investigate the AHE in thin bcc Fe(001) films of 3-, 5-, 7-
and 11-layer thickness. This approximately corresponds to the thickness of $0.6$, $0.9$, $1.2$ and
2~nm, respectively. To our knowledge, no experimental data exists for such ultra-thin Fe films. As can be 
seen from our calculations, presented in Fig.~\ref{fig:film}, the behavior of the AHE for bcc Fe in the limit 
of ultra-thin films is very different from that in bulk, given in Fig.~\ref{fig:1}. The side-jump contribution
in 3-layer film is very small, but it rises in magnitude significantly as the film thickness is increased, reaching as much as $-$700~S/cm for 11 layers, and exhibiting a change of sign at small values of
$\rho_{xx}$. Increasing thickness further towards the bulk limit will bring the magnitude of $\sigma^{sj}$
down, and the change-of-sign point to higher values of $\rho_{xx}$, see Fig.~\ref{fig:1}. The intrinsic
skew-scattering contributes mostly in 3- and 5-layer films, while its magnitude gets significantly smaller as the thickness is increased, see also Fig.~\ref{fig:1}. Remarkably, for 3- and 5-layer films $\sigma^{isk}$ hits as 
much as 1000~S/cm in magnitude, and its dependence on $\rho_{xx}$ determines the behavior of overall AHC, which
even exhibits a change of sign for 3 layers. On the other hand the intrinsic part of the AHC determines
the total AHE with increasing thickness, both in magnitude and behavior, in consistency with the calculations for bulk Fe. And while the 3-layer $\sigma^{int}$ does not depend on $\rho_{xx}$, this
dependence is quantitatively consistent with that in bulk Fe for larger thickness, although it is much more pronounced.
Overall, one has to point out that the intrinsic values in clean limit are rather far off the bulk value, except for the 5-layer film. The insets in Fig.~\ref{fig:film} show that unlike in most of the
bulk ferromagnets considered previously the behavior of $\rho_{yx}/\rho_{xx}$ is strongly non-linear, which indicates that the common scaling laws which predict linearity do not hold in ultra-thin films.

To conclude, in the limit of ultra-thin films one can expect large changes in the total value of the AHE,
its sign, and the magnitude of the contributions of different origin. This marks the few-layer thin nanostructures
of ferromagnets as an exciting type of systems to study in the future, both experimentally and theoretically.

\section{Conclusions}

In summary, we have implemented a method for calculating the AHC in disordered ferromagnets solely from the electronic structure of the perfect crystal. We used Gaussian model for disorder potential that allows us to tune disorder via a single parameter, presenting thus a significant simplification as compared to more elaborated schemes. In the clean limit, within
our approach we arrive at all scattering-independent contributions to the AHE: intrinsic, side jump and intrinsic skew-scattering
contributions. In particular the latter has been never evaluated in real materials. 

We implemented our method within the 
FLAPW code {\tt FLEUR} and applied it to a number of bulk ferromagnets and thin films. 
We found that in most materials our model is able to reproduce the correct qualitative behavior of
$\rho_{yx}/\rho_{xx}$ as a function of the longitudinal resistivity, generally providing also rather good quantitative agreement with experiments.
In particular, we were able to describe the temperature dependence of the AHE from the knowledge of the experimental $\rho_{xx}$. Within our approach we transparently separated the intrinsic, side jump, and intrinsic skew scattering contributions to the AHE and studied their respective resistivity/temperature dependences, finding that in most cases they all exhibit a non-trivial behavior. In particular we demonstrated that in most ferromagnets the intrinsic skew scattering improves agreement with experiments in that it provides a significant contribution that can even exceed the intrinsic and the side jump contributions at low resistivities.

\section{acknowledgments}

This work was financially supported by SPP 1538 SpinCaT programme of the Deutsche Forschungsgemeinschaft,
and the HGF YIG VH-NG-513 project of the Helmholtz Gemeinschaft.
We would like to thank Xiaofeng Jin, Diemo K\"odderitzsch and Jairo Sinova for fruitful discussions.
We are grateful to J\"ulich Supercomputing Centre for providing us with computational time.


\begin{thebibliography}{29}
\expandafter\ifx\csname natexlab\endcsname\relax\def\natexlab#1{#1}\fi
\expandafter\ifx\csname bibnamefont\endcsname\relax
  \def\bibnamefont#1{#1}\fi
\expandafter\ifx\csname bibfnamefont\endcsname\relax
  \def\bibfnamefont#1{#1}\fi
\expandafter\ifx\csname citenamefont\endcsname\relax
  \def\citenamefont#1{#1}\fi
\expandafter\ifx\csname url\endcsname\relax
  \def\url#1{\texttt{#1}}\fi
\expandafter\ifx\csname urlprefix\endcsname\relax\def\urlprefix{URL }\fi
\providecommand{\bibinfo}[2]{#2}
\providecommand{\eprint}[2][]{\url{#2}}

\bibitem[{\citenamefont{Hall}(1881)}]{Hall}
\bibinfo{author}{\bibfnamefont{E.}~\bibnamefont{Hall}},
  \bibinfo{journal}{Philosophical Magazine Series 5}
  \textbf{\bibinfo{volume}{12}}, \bibinfo{pages}{157} (\bibinfo{year}{1881}),
  \urlprefix\url{http://www.tandfonline.com/doi/abs/10.1080/14786448108627086}.

\bibitem[{\citenamefont{Nagaosa}(2006)}]{Nagaosa06}
\bibinfo{author}{\bibfnamefont{N.}~\bibnamefont{Nagaosa}},
  \bibinfo{journal}{Journal of the Physical Society of Japan}
  \textbf{\bibinfo{volume}{75}}, \bibinfo{pages}{042001}
  (\bibinfo{year}{2006}),
  \urlprefix\url{http://jpsj.ipap.jp/link?JPSJ/75/042001/}.

\bibitem[{\citenamefont{Karplus and Luttinger}(1954)}]{Karplus}
\bibinfo{author}{\bibfnamefont{R.}~\bibnamefont{Karplus}} \bibnamefont{and}
  \bibinfo{author}{\bibfnamefont{J.~M.} \bibnamefont{Luttinger}},
  \bibinfo{journal}{Phys. Rev.} \textbf{\bibinfo{volume}{95}},
  \bibinfo{pages}{1154} (\bibinfo{year}{1954}),
  \urlprefix\url{http://link.aps.org/doi/10.1103/PhysRev.95.1154}.

\bibitem[{\citenamefont{Nagaosa et~al.}(2010)\citenamefont{Nagaosa, Sinova,
  Onoda, MacDonald, and Ong}}]{Nagaosa}
\bibinfo{author}{\bibfnamefont{N.}~\bibnamefont{Nagaosa}},
  \bibinfo{author}{\bibfnamefont{J.}~\bibnamefont{Sinova}},
  \bibinfo{author}{\bibfnamefont{S.}~\bibnamefont{Onoda}},
  \bibinfo{author}{\bibfnamefont{A.~H.} \bibnamefont{MacDonald}},
  \bibnamefont{and} \bibinfo{author}{\bibfnamefont{N.~P.} \bibnamefont{Ong}},
  \bibinfo{journal}{Rev. Mod. Phys.} \textbf{\bibinfo{volume}{82}},
  \bibinfo{pages}{1539} (\bibinfo{year}{2010}),
  \urlprefix\url{http://link.aps.org/doi/10.1103/RevModPhys.82.1539}.

\bibitem[{\citenamefont{Smit}(1958)}]{Smit58}
\bibinfo{author}{\bibfnamefont{J.}~\bibnamefont{Smit}},
  \bibinfo{journal}{Physica} \textbf{\bibinfo{volume}{24}}, \bibinfo{pages}{39
  } (\bibinfo{year}{1958}), ISSN \bibinfo{issn}{0031-8914},
  \urlprefix\url{http://www.sciencedirect.com/science/article/pii/S00318914589%
35419}.

\bibitem[{\citenamefont{Berger}(1970)}]{Berger70}
\bibinfo{author}{\bibfnamefont{L.}~\bibnamefont{Berger}},
  \bibinfo{journal}{Phys. Rev. B} \textbf{\bibinfo{volume}{2}},
  \bibinfo{pages}{4559} (\bibinfo{year}{1970}),
  \urlprefix\url{http://link.aps.org/doi/10.1103/PhysRevB.2.4559}.

\bibitem[{\citenamefont{Dheer}(1967)}]{Dheer}
\bibinfo{author}{\bibfnamefont{P.~N.} \bibnamefont{Dheer}},
  \bibinfo{journal}{Phys. Rev.} \textbf{\bibinfo{volume}{156}},
  \bibinfo{pages}{637} (\bibinfo{year}{1967}),
  \urlprefix\url{http://link.aps.org/doi/10.1103/PhysRev.156.637}.

\bibitem[{\citenamefont{Smit}(1955)}]{Smit55}
\bibinfo{author}{\bibfnamefont{J.}~\bibnamefont{Smit}},
  \bibinfo{journal}{Physica} \textbf{\bibinfo{volume}{21}}, \bibinfo{pages}{877
  } (\bibinfo{year}{1955}), ISSN \bibinfo{issn}{0031-8914},
  \urlprefix\url{http://www.sciencedirect.com/science/article/pii/S00318914559%
25969}.

\bibitem[{\citenamefont{K\"otzler and Gil}(2005)}]{Koetzler}
\bibinfo{author}{\bibfnamefont{J.}~\bibnamefont{K\"otzler}} \bibnamefont{and}
  \bibinfo{author}{\bibfnamefont{W.}~\bibnamefont{Gil}},
  \bibinfo{journal}{Phys. Rev. B} \textbf{\bibinfo{volume}{72}},
  \bibinfo{pages}{060412} (\bibinfo{year}{2005}),
  \urlprefix\url{http://link.aps.org/doi/10.1103/PhysRevB.72.060412}.

\bibitem[{\citenamefont{Tian et~al.}(2009)\citenamefont{Tian, Ye, and
  Jin}}]{Tian}
\bibinfo{author}{\bibfnamefont{Y.}~\bibnamefont{Tian}},
  \bibinfo{author}{\bibfnamefont{L.}~\bibnamefont{Ye}}, \bibnamefont{and}
  \bibinfo{author}{\bibfnamefont{X.}~\bibnamefont{Jin}},
  \bibinfo{journal}{Phys. Rev. Lett.} \textbf{\bibinfo{volume}{103}},
  \bibinfo{pages}{087206} (\bibinfo{year}{2009}),
  \urlprefix\url{http://link.aps.org/doi/10.1103/PhysRevLett.103.087206}.

\bibitem[{\citenamefont{Ye et~al.}(2012)\citenamefont{Ye, Tian, Jin, and
  Xiao}}]{Ye}
\bibinfo{author}{\bibfnamefont{L.}~\bibnamefont{Ye}},
  \bibinfo{author}{\bibfnamefont{Y.}~\bibnamefont{Tian}},
  \bibinfo{author}{\bibfnamefont{X.}~\bibnamefont{Jin}}, \bibnamefont{and}
  \bibinfo{author}{\bibfnamefont{D.}~\bibnamefont{Xiao}},
  \bibinfo{journal}{Phys. Rev. B} \textbf{\bibinfo{volume}{85}},
  \bibinfo{pages}{220403} (\bibinfo{year}{2012}),
  \urlprefix\url{http://link.aps.org/doi/10.1103/PhysRevB.85.220403}.

\bibitem[{\citenamefont{Lowitzer et~al.}(2010)\citenamefont{Lowitzer,
  K\"odderitzsch, and Ebert}}]{Lowitzer}
\bibinfo{author}{\bibfnamefont{S.}~\bibnamefont{Lowitzer}},
  \bibinfo{author}{\bibfnamefont{D.}~\bibnamefont{K\"odderitzsch}},
  \bibnamefont{and} \bibinfo{author}{\bibfnamefont{H.}~\bibnamefont{Ebert}},
  \bibinfo{journal}{Phys. Rev. Lett.} \textbf{\bibinfo{volume}{105}},
  \bibinfo{pages}{266604} (\bibinfo{year}{2010}),
  \urlprefix\url{http://link.aps.org/doi/10.1103/PhysRevLett.105.266604}.

\bibitem[{\citenamefont{Turek et~al.}(2012)\citenamefont{Turek, Kudrnovsk\'y,
  and Drchal}}]{Turek}
\bibinfo{author}{\bibfnamefont{I.}~\bibnamefont{Turek}},
  \bibinfo{author}{\bibfnamefont{J.}~\bibnamefont{Kudrnovsk\'y}},
  \bibnamefont{and} \bibinfo{author}{\bibfnamefont{V.}~\bibnamefont{Drchal}},
  \bibinfo{journal}{Phys. Rev. B} \textbf{\bibinfo{volume}{86}},
  \bibinfo{pages}{014405} (\bibinfo{year}{2012}),
  \urlprefix\url{http://link.aps.org/doi/10.1103/PhysRevB.86.014405}.

\bibitem[{\citenamefont{K\"odderitzsch
  et~al.}(2013)\citenamefont{K\"odderitzsch, Chadova, Minar, and
  Ebert}}]{Koedderitzsch}
\bibinfo{author}{\bibfnamefont{D.}~\bibnamefont{K\"odderitzsch}},
  \bibinfo{author}{\bibfnamefont{K.}~\bibnamefont{Chadova}},
  \bibinfo{author}{\bibfnamefont{J.}~\bibnamefont{Minar}}, \bibnamefont{and}
  \bibinfo{author}{\bibfnamefont{H.}~\bibnamefont{Ebert}},
  \bibinfo{journal}{New Journal of Physics} \textbf{\bibinfo{volume}{15}},
  \bibinfo{pages}{053009} (\bibinfo{year}{2013}),
  \urlprefix\url{http://stacks.iop.org/1367-2630/15/i=5/a=053009}.

\bibitem[{\citenamefont{Kovalev et~al.}(2010)\citenamefont{Kovalev, Sinova, and
  Tserkovnyak}}]{Kovalev}
\bibinfo{author}{\bibfnamefont{A.~A.} \bibnamefont{Kovalev}},
  \bibinfo{author}{\bibfnamefont{J.}~\bibnamefont{Sinova}}, \bibnamefont{and}
  \bibinfo{author}{\bibfnamefont{Y.}~\bibnamefont{Tserkovnyak}},
  \bibinfo{journal}{Phys. Rev. Lett.} \textbf{\bibinfo{volume}{105}},
  \bibinfo{pages}{036601} (\bibinfo{year}{2010}),
  \urlprefix\url{http://link.aps.org/doi/10.1103/PhysRevLett.105.036601}.

\bibitem[{\citenamefont{Weischenberg et~al.}(2011)\citenamefont{Weischenberg,
  Freimuth, Sinova, Bl\"ugel, and Mokrousov}}]{Weischenberg}
\bibinfo{author}{\bibfnamefont{J.}~\bibnamefont{Weischenberg}},
  \bibinfo{author}{\bibfnamefont{F.}~\bibnamefont{Freimuth}},
  \bibinfo{author}{\bibfnamefont{J.}~\bibnamefont{Sinova}},
  \bibinfo{author}{\bibfnamefont{S.}~\bibnamefont{Bl\"ugel}}, \bibnamefont{and}
  \bibinfo{author}{\bibfnamefont{Y.}~\bibnamefont{Mokrousov}},
  \bibinfo{journal}{Phys. Rev. Lett.} \textbf{\bibinfo{volume}{107}},
  \bibinfo{pages}{106601} (\bibinfo{year}{2011}),
  \urlprefix\url{http://link.aps.org/doi/10.1103/PhysRevLett.107.106601}.

\bibitem[{\citenamefont{Sinitsyn et~al.}(2007)\citenamefont{Sinitsyn,
  MacDonald, Jungwirth, Dugaev, and Sinova}}]{Sinitsyn-2}
\bibinfo{author}{\bibfnamefont{N.~A.} \bibnamefont{Sinitsyn}},
  \bibinfo{author}{\bibfnamefont{A.~H.} \bibnamefont{MacDonald}},
  \bibinfo{author}{\bibfnamefont{T.}~\bibnamefont{Jungwirth}},
  \bibinfo{author}{\bibfnamefont{V.~K.} \bibnamefont{Dugaev}},
  \bibnamefont{and} \bibinfo{author}{\bibfnamefont{J.}~\bibnamefont{Sinova}},
  \bibinfo{journal}{Phys. Rev. B} \textbf{\bibinfo{volume}{75}},
  \bibinfo{pages}{045315} (\bibinfo{year}{2007}),
  \urlprefix\url{http://link.aps.org/doi/10.1103/PhysRevB.75.045315}.

\bibitem[{\citenamefont{Sinitsyn}(2008)}]{Sinitsyn}
\bibinfo{author}{\bibfnamefont{N.~A.} \bibnamefont{Sinitsyn}},
  \bibinfo{journal}{Journal of Physics: Condensed Matter}
  \textbf{\bibinfo{volume}{20}}, \bibinfo{pages}{023201}
  (\bibinfo{year}{2008}),
  \urlprefix\url{http://stacks.iop.org/0953-8984/20/i=2/a=023201}.

\bibitem[{FLE()}]{FLEUR}
\urlprefix\url{http://www.flapw.de}.

\bibitem[{\citenamefont{Souza et~al.}(2001)\citenamefont{Souza, Marzari, and
  Vanderbilt}}]{Souza01}
\bibinfo{author}{\bibfnamefont{I.}~\bibnamefont{Souza}},
  \bibinfo{author}{\bibfnamefont{N.}~\bibnamefont{Marzari}}, \bibnamefont{and}
  \bibinfo{author}{\bibfnamefont{D.}~\bibnamefont{Vanderbilt}},
  \bibinfo{journal}{Phys. Rev. B} \textbf{\bibinfo{volume}{65}},
  \bibinfo{pages}{035109} (\bibinfo{year}{2001}),
  \urlprefix\url{http://link.aps.org/doi/10.1103/PhysRevB.65.035109}.

\bibitem[{\citenamefont{Wildberger et~al.}(1995)\citenamefont{Wildberger, Lang,
  Zeller, and Dederichs}}]{Wildberger}
\bibinfo{author}{\bibfnamefont{K.}~\bibnamefont{Wildberger}},
  \bibinfo{author}{\bibfnamefont{P.}~\bibnamefont{Lang}},
  \bibinfo{author}{\bibfnamefont{R.}~\bibnamefont{Zeller}}, \bibnamefont{and}
  \bibinfo{author}{\bibfnamefont{P.~H.} \bibnamefont{Dederichs}},
  \bibinfo{journal}{Phys. Rev. B} \textbf{\bibinfo{volume}{52}},
  \bibinfo{pages}{11502} (\bibinfo{year}{1995}),
  \urlprefix\url{http://link.aps.org/doi/10.1103/PhysRevB.52.11502}.

\bibitem[{\citenamefont{Mostofi et~al.}(2008)\citenamefont{Mostofi, Yates, Lee,
  Souza, Vanderbilt, and Marzari}}]{wannier90}
\bibinfo{author}{\bibfnamefont{A.~A.} \bibnamefont{Mostofi}},
  \bibinfo{author}{\bibfnamefont{J.~R.} \bibnamefont{Yates}},
  \bibinfo{author}{\bibfnamefont{Y.-S.} \bibnamefont{Lee}},
  \bibinfo{author}{\bibfnamefont{I.}~\bibnamefont{Souza}},
  \bibinfo{author}{\bibfnamefont{D.}~\bibnamefont{Vanderbilt}},
  \bibnamefont{and} \bibinfo{author}{\bibfnamefont{N.}~\bibnamefont{Marzari}},
  \bibinfo{journal}{Computer Physics Communications}
  \textbf{\bibinfo{volume}{178}}, \bibinfo{pages}{685 } (\bibinfo{year}{2008}),
  ISSN \bibinfo{issn}{0010-4655},
  \urlprefix\url{http://www.sciencedirect.com/science/article/pii/S00104655070%
04936}.

\bibitem[{\citenamefont{Freimuth et~al.}(2008)\citenamefont{Freimuth,
  Mokrousov, Wortmann, Heinze, and Bl\"ugel}}]{interface}
\bibinfo{author}{\bibfnamefont{F.}~\bibnamefont{Freimuth}},
  \bibinfo{author}{\bibfnamefont{Y.}~\bibnamefont{Mokrousov}},
  \bibinfo{author}{\bibfnamefont{D.}~\bibnamefont{Wortmann}},
  \bibinfo{author}{\bibfnamefont{S.}~\bibnamefont{Heinze}}, \bibnamefont{and}
  \bibinfo{author}{\bibfnamefont{S.}~\bibnamefont{Bl\"ugel}},
  \bibinfo{journal}{Phys. Rev. B} \textbf{\bibinfo{volume}{78}},
  \bibinfo{pages}{035120} (\bibinfo{year}{2008}),
  \urlprefix\url{http://link.aps.org/doi/10.1103/PhysRevB.78.035120}.

\bibitem[{\citenamefont{Hou et~al.}(2012)\citenamefont{Hou, Li, Wei, Tian, Wu,
  and Jin}}]{Hou}
\bibinfo{author}{\bibfnamefont{D.}~\bibnamefont{Hou}},
  \bibinfo{author}{\bibfnamefont{Y.}~\bibnamefont{Li}},
  \bibinfo{author}{\bibfnamefont{D.}~\bibnamefont{Wei}},
  \bibinfo{author}{\bibfnamefont{D.}~\bibnamefont{Tian}},
  \bibinfo{author}{\bibfnamefont{L.}~\bibnamefont{Wu}}, \bibnamefont{and}
  \bibinfo{author}{\bibfnamefont{X.}~\bibnamefont{Jin}},
  \bibinfo{journal}{Journal of Physics: Condensed Matter}
  \textbf{\bibinfo{volume}{24}}, \bibinfo{pages}{482001}
  (\bibinfo{year}{2012}),
  \urlprefix\url{http://stacks.iop.org/0953-8984/24/i=48/a=482001}.

\bibitem[{\citenamefont{Seemann
  et~al.}(2010{\natexlab{a}})\citenamefont{Seemann, Mokrousov, Aziz, Miguel,
  Kronast, Kuch, Blamire, Hindmarch, Hickey, Souza et~al.}}]{Seemann}
\bibinfo{author}{\bibfnamefont{K.~M.} \bibnamefont{Seemann}},
  \bibinfo{author}{\bibfnamefont{Y.}~\bibnamefont{Mokrousov}},
  \bibinfo{author}{\bibfnamefont{A.}~\bibnamefont{Aziz}},
  \bibinfo{author}{\bibfnamefont{J.}~\bibnamefont{Miguel}},
  \bibinfo{author}{\bibfnamefont{F.}~\bibnamefont{Kronast}},
  \bibinfo{author}{\bibfnamefont{W.}~\bibnamefont{Kuch}},
  \bibinfo{author}{\bibfnamefont{M.~G.} \bibnamefont{Blamire}},
  \bibinfo{author}{\bibfnamefont{A.~T.} \bibnamefont{Hindmarch}},
  \bibinfo{author}{\bibfnamefont{B.~J.} \bibnamefont{Hickey}},
  \bibinfo{author}{\bibfnamefont{I.}~\bibnamefont{Souza}},
  \bibnamefont{et~al.}, \bibinfo{journal}{Phys. Rev. Lett.}
  \textbf{\bibinfo{volume}{104}}, \bibinfo{pages}{076402}
  (\bibinfo{year}{2010}{\natexlab{a}}),
  \urlprefix\url{http://link.aps.org/doi/10.1103/PhysRevLett.104.076402}.

\bibitem[{\citenamefont{He et~al.}(2012)\citenamefont{He, Ma, Shi, Guo, Zheng,
  Xin, and Zhou}}]{He}
\bibinfo{author}{\bibfnamefont{P.}~\bibnamefont{He}},
  \bibinfo{author}{\bibfnamefont{L.}~\bibnamefont{Ma}},
  \bibinfo{author}{\bibfnamefont{Z.}~\bibnamefont{Shi}},
  \bibinfo{author}{\bibfnamefont{G.~Y.} \bibnamefont{Guo}},
  \bibinfo{author}{\bibfnamefont{J.-G.} \bibnamefont{Zheng}},
  \bibinfo{author}{\bibfnamefont{Y.}~\bibnamefont{Xin}}, \bibnamefont{and}
  \bibinfo{author}{\bibfnamefont{S.~M.} \bibnamefont{Zhou}},
  \bibinfo{journal}{Phys. Rev. Lett.} \textbf{\bibinfo{volume}{109}},
  \bibinfo{pages}{066402} (\bibinfo{year}{2012}),
  \urlprefix\url{http://link.aps.org/doi/10.1103/PhysRevLett.109.066402}.

\bibitem[{\citenamefont{Ho et~al.}(1983)\citenamefont{Ho, Ackerman, Wu, Havill,
  Bogaard, Matula, Oh, and James}}]{Ho}
\bibinfo{author}{\bibfnamefont{C.~Y.} \bibnamefont{Ho}},
  \bibinfo{author}{\bibfnamefont{M.~W.} \bibnamefont{Ackerman}},
  \bibinfo{author}{\bibfnamefont{K.~Y.} \bibnamefont{Wu}},
  \bibinfo{author}{\bibfnamefont{T.~N.} \bibnamefont{Havill}},
  \bibinfo{author}{\bibfnamefont{R.~H.} \bibnamefont{Bogaard}},
  \bibinfo{author}{\bibfnamefont{R.~A.} \bibnamefont{Matula}},
  \bibinfo{author}{\bibfnamefont{S.~G.} \bibnamefont{Oh}}, \bibnamefont{and}
  \bibinfo{author}{\bibfnamefont{H.~M.} \bibnamefont{James}},
  \bibinfo{journal}{Journal of Physical and Chemical Reference Data}
  \textbf{\bibinfo{volume}{12}}, \bibinfo{pages}{183} (\bibinfo{year}{1983}),
  \urlprefix\url{http://link.aip.org/link/?JPR/12/183/1}.

\bibitem[{\citenamefont{Roman et~al.}(2009)\citenamefont{Roman, Mokrousov, and
  Souza}}]{Roman}
\bibinfo{author}{\bibfnamefont{E.}~\bibnamefont{Roman}},
  \bibinfo{author}{\bibfnamefont{Y.}~\bibnamefont{Mokrousov}},
  \bibnamefont{and} \bibinfo{author}{\bibfnamefont{I.}~\bibnamefont{Souza}},
  \bibinfo{journal}{Phys. Rev. Lett.} \textbf{\bibinfo{volume}{103}},
  \bibinfo{pages}{097203} (\bibinfo{year}{2009}),
  \urlprefix\url{http://link.aps.org/doi/10.1103/PhysRevLett.103.097203}.

\bibitem[{\citenamefont{Seemann
  et~al.}(2010{\natexlab{b}})\citenamefont{Seemann, Hickey, Baltz, Hickey, and
  Marrows}}]{Seemann-NJP}
\bibinfo{author}{\bibfnamefont{K.~M.} \bibnamefont{Seemann}},
  \bibinfo{author}{\bibfnamefont{M.~C.} \bibnamefont{Hickey}},
  \bibinfo{author}{\bibfnamefont{V.}~\bibnamefont{Baltz}},
  \bibinfo{author}{\bibfnamefont{B.~J.} \bibnamefont{Hickey}},
  \bibnamefont{and} \bibinfo{author}{\bibfnamefont{C.~H.}
  \bibnamefont{Marrows}}, \bibinfo{journal}{New Journal of Physics}
  \textbf{\bibinfo{volume}{12}}, \bibinfo{pages}{033033}
  (\bibinfo{year}{2010}{\natexlab{b}}),
  \urlprefix\url{http://stacks.iop.org/1367-2630/12/i=3/a=033033}.

\end{thebibliography}
\end{document}